\documentclass[twocolumn,aps,prd,preprintnumbers,superscriptaddress,nofootinbib,amsmath,amssymb,floats,floatfix,showkeys,notitlepage,showpacs]{revtex4-1}
\usepackage{orcidlink}
\usepackage{comment}
\usepackage{lipsum}
\usepackage{graphicx}
\usepackage{subfigure}
\usepackage{palatino}
\usepackage{sans}
\usepackage{hyperref}
\hypersetup{colorlinks=true,linkcolor=blue,urlcolor=blue,citecolor=blue}
\usepackage[toc,page]{appendix}
\usepackage[normalem]{ulem}
\usepackage{adjustbox}
\usepackage{latexsym}
\usepackage{amsmath}
\usepackage{amssymb}
\usepackage{amsfonts}
\usepackage{dcolumn}
\usepackage{bm}
\usepackage{tikz}
\usepackage{bigints}
\usepackage{array,tabularx,multirow,booktabs}
\usepackage[tracking=true]{microtype}
\usepackage{multirow}
\SetTracking{}{500}
\SetTracking{encoding={*}, shape=sc}{40}
\UseRawInputEncoding %for inputenc error%
\allowdisplaybreaks
 \newcommand{\bq}{\begin{equation}}
 \newcommand{\eq}{\end{equation}}
 \newcommand{\bqn}{\begin{equation}}
 \newcommand{\eqn}{\end{equation}}
 
 \newcommand{\lb}{\label}

\newcommand{\be}{\nopagebreak[3]\begin{equation}}
\newcommand{\ee}{\end{equation}}

\newcommand{\ba}{\nopagebreak[3]\begin{equation}}
\newcommand{\ea}{\end{equation}}

\NewDocumentCommand{\evalat}{sO{\big}mm}{%
  \IfBooleanTF{#1}
   {\mleft. #3 \mright|_{#4}}
   {#3#2|_{#4}}%
}

\begin{document} \sloppy
	%%%%%%%%%%%%%%%%%%%%%%%%%%%%%%%%%%%%%%%%%%%%%%%%%%%%%%%%%%%%%%%%%%%%%%%%%%%%%%%%%%%%%%
	\newcommand \nn{\nonumber}
	\newcommand \fc{\frac}
	\newcommand \lt{\left}
	\newcommand \rt{\right}
	\newcommand \pd{\partial}
	\newcommand \e{\text{e}}
	\newcommand \hmn{h_{\mu\nu}}
	
	\newcommand{\PC}[1]{\ensuremath{\left(#1\right)}} % parenteses curvos do tamanho adequado
	\newcommand{\PX}[1]{\ensuremath{\left\lbrace#1\right\rbrace}} % chavetas do tamanho adequado
	\newcommand{\BR}[1]{\ensuremath{\left\langle#1\right\vert}} % Bra do tamanho adequado
	\newcommand{\KT}[1]{\ensuremath{\left\vert#1\right\rangle}} % Ket do tamanho adequado
	\newcommand{\MD}[1]{\ensuremath{\left\vert#1\right\vert}} % modulo do tamanho adequado

\title{Black hole in the Dekel-Zhao dark matter profile}

\author{Ali \"Ovg\"un \orcidlink{0000-0002-9889-342X}}
\email{ali.ovgun@emu.edu.tr}
\affiliation{Physics Department, Eastern Mediterranean University, Famagusta, 99628 North
Cyprus via Mersin 10, Turkiye.}

\author{Reggie C. Pantig \orcidlink{0000-0002-3101-8591}}
\email{rcpantig@mapua.edu.ph}
\affiliation{Physics Department, Map\'ua University, 658 Muralla St., Intramuros, Manila 1002, Philippines}

\begin{abstract}
Motivated by the work of Cardoso et al. [Phys. Rev. D 105 (2022) 6, L061501, https://doi.org/10.1103/PhysRevD.105.L061501] on black holes in galaxies, we derive a new black hole solution surrounded by a Dekel-Zhao (DZ) dark matter profile. The derived metric, influenced by DZ profile parameters, exhibits two distinct regimes: for $r \ll r_{\rm c}$ (\textcolor{black}{$r_{\rm c}$ is a characteristic radius}), exponential corrections dominate, producing significant deviations from the Schwarzschild solution near dense cores, while for $r \gg r_{\rm c}$, these corrections vanish, restoring the Schwarzschild metric at large distances. These findings ensure consistency with general relativity in vacuum. \textcolor{black}{The black hole shadow and deflection angle are analyzed, demonstrating that the shadow radius increases with black hole mass ($M_{\rm BH}$), while higher central densities ($\rho_{\rm ch}$) result in smaller shadows, reflecting the environmental impact of dense dark matter halos.} Photon dynamics reveal how DZ profiles modify critical impact parameters and effective potentials, with gravitational lensing effects highly sensitive to the characteristic radius ($r_{\rm c}$). Smaller $r_{\rm c}$ values lead to larger deflection angles due to stronger gravitational effects near compact cores. This work highlights the significance dark matter profiles in shaping black hole observables, providing a theoretical foundation for future observational studies and advancing the understanding of dark matter-black hole interactions in astrophysical and cosmological contexts.
\end{abstract}	



\maketitle

%\tableofcontents

\section{Introduction}\label{intro}
%Black holes
Black hole physics is a cornerstone of modern theoretical and observational astrophysics, offering profound insights into the nature of gravity, spacetime, and fundamental physics. Black holes, predicted by Einstein's general theory of relativity, represent regions of spacetime where gravitational forces are so intense that not even light can escape. They serve as natural laboratories for exploring the limits of classical and quantum theories, providing a unique setting where extreme conditions challenge our understanding of fundamental forces. From an observational perspective, black holes are central to the study of high-energy astrophysical phenomena, such as accretion processes, relativistic jets, and gamma-ray bursts. Supermassive black holes, found at the centers of galaxies, play a crucial role in galaxy formation and evolution, influencing their host environments through feedback mechanisms. Stellar-mass black holes, on the other hand, offer insights into the final stages of stellar evolution and the dynamics of compact objects.

%Dark matter
Dark matter, one of the most enigmatic components of the universe, has captured the imagination of scientists for decades due to its profound implications for both cosmology and particle physics. Its significance in research cannot be overstated, as it accounts for approximately 26.8\% of the universe's total mass-energy content, influencing galaxy formation, the cosmic microwave background, and the large-scale structure of the cosmos. The quest to understand dark matter is not merely about confirming its existence, which is well-supported by gravitational effects observed in galaxy rotation curves, gravitational lensing, and the dynamics of galaxy clusters \cite{DarkMatter,Marsh:2024ury,DelNobile:2021wmp}, but also about deciphering its nature. This endeavor potentially opens doors to physics beyond the Standard Model, challenging our fundamental understanding of particle interactions and the forces that govern the universe. 

Historically, attempts to directly detect dark matter particles, particularly Weakly Interacting Massive Particles (WIMPs), have largely been met with null results, casting doubt on some of the most favored candidates. Experiments like the Cryogenic Dark Matter Search (CDMS) and the DAMA/LIBRA have claimed potential signals, but these have been contentious, leading to a complex tapestry of both success and failure in detection efforts \cite{Bernabei:2018jrt,PICO:2017tgi,CRESST:2015txj,Bernabei:2013xsa,DAMA:2008jlt}. Today, the field is witnessing a renaissance with advanced experiments such as LUX-ZEPLIN (LZ) and XENONnT, which aim to push the sensitivity boundaries to unprecedented levels \cite{LZ:2022lsv}. These experiments utilize large quantities of liquid xenon to detect the faint signals of dark matter interactions, with LZ targeting a sensitivity to spin-independent WIMP-nucleon cross-sections down to $10^{-48} \text{ cm}^2$ before the neutrino background becomes significant. Alongside direct detection, there are also ongoing indirect searches for dark matter annihilation signals and efforts at particle colliders like the LHC to produce dark matter particles \cite{Abdallah:2015ter,Askew:2014kqa}. Even using the Earth itself, can serve as a dark matter detector \cite{Baum:2018tfw}. These efforts reflect a multi-faceted approach to unraveling the cosmic mystery of dark matter.

%Connecting dark matter to black hole phyiscs
The idea of indirectly detecting dark matter through its interactions with extreme objects like astrophysical black holes offers a fascinating avenue in our quest to understand this elusive component of the universe. Black holes, with their intense gravitational fields, could serve as cosmic laboratories where dark matter particles might either accumulate or annihilate, producing detectable signals, which serve to develop certain methods on these paradigms \cite{Xu:2018wow,Konoplya:2022hbl,Cardoso:2021wlq,Gomez:2024ack}.  Exploring dark matter detection via the shadow cast by black holes and the weak deflection angle of light offers a unique approach leveraging these cosmic giants' gravitational environments. The shadow of a black hole, as seen in observations by the Event Horizon Telescope, could be altered by dark matter halos, potentially showing a larger or differently shaped shadow than expected by general relativity alone. Similarly, dark matter near black holes might enhance gravitational lensing, subtly changing light paths from background sources, which could be detected as anomalies in lensing patterns. These methods not only probe gravity in extreme conditions but also provide indirect evidence for dark matter, akin to deciphering invisible ink on a dark canvas with the right illumination. Several studies have explored this direction \cite{Maeda:2024tsg,Shen:2024qxv,Qiao:2024ehj,Wu:2024hxr,Capozziello:2023tbo,Capozziello:2023rfv,Liu:2023xtb,Xu:2021dkv,Xu:2020jpv,Jusufi:2020cpn,Haroon:2018ryd,Hou:2018bar,Hou:2018avu,Pantig:2024rmr,Ovgun:2023wmc,Pantig:2022sjb,Biswas:2023ofz,Feng:2022evy,Chakraborty:2024gcr,Rahman:2023sof}.

There are many dark matter models present in the literature. In this paper, we are interested in the Dekel-Zhao profile, which offers a versatile framework for modeling dark matter halos, enabling researchers to capture the diverse structural properties observed in astrophysical systems \cite{Zhao:1995cp,Zhao:1996mr} . This double power-law density profile transitions smoothly between different slopes at small and large radii, making it highly adaptable to both observational data and theoretical predictions. It is defined as \cite{Zhao:1995cp,Zhao:1996mr} :
\begin{equation} \label{dmprofile1}
\rho(r) = \frac{\rho_{\rm ch}}{x^a \left(1 + x^{1/b}\right)^{b(g-a)}},
\end{equation}
\textcolor{black}{where $x = r / r_{\rm c}$ is the radial distance normalized to a characteristic radius $r_{\rm c}$,} and $\rho_{\rm ch}$ represents a characteristic density. The parameters $a$, $b$, and $g$ govern the inner slope, transition sharpness, and outer slope of the profile, respectively. By adjusting these parameters, the Dekel-Zhao profile can replicate a wide range of observed dark matter distributions, providing critical insights into the properties and dynamics of dark matter in the universe.

%Motivation
Our aim in this work is to investigate the impact of the Dekel-Zhao (DZ) dark matter profile on black hole formation, utilizing a generalized version of the "Einstein cluster" method as presented in \cite{Cardoso:2021wlq} to derive the black hole metric. Such a method has been used by several authors to contribute to the literature of dark matter detection in gravitationally extreme conditions \cite{Xamidov:2025hrj,Chen:2024lpd,Gohain:2024eer,Macedo:2024qky,Montalvo:2024iwq,Zhang:2024ugv,Pezzella:2024tkf,Mollicone:2024lxy,Destounis:2022obl,Figueiredo:2023gas,Speeney:2024mas,Polcar:2022bwv,Chen:2023akf}.

We organize this paper as follows: In Sect. \ref{sec2}, we derive new spherically symmetric black hole solutions surrounded by Dekel-Zhao (DZ) profile halos. In Sect. \ref{sec3}, we analyze the shadow of the black hole. Finally, Sect. \ref{conc} provides a summary of our findings and presents some recommendations for future work.

\section{Exact solution of black holes surrounded by DZ profile halos using method of Cardoso et al. }
 \label{sec2}
To obtain an exact field solution for a black hole immersed in dark matter, Cardoso et al. \cite{Cardoso:2021wlq} utilized a generalized version of the "Einstein cluster" method. In this paper, we use this approach approach to derive new black hole solutions surrounded by DZ profile. This method is applied in many  different cases \cite{Konoplya:2022hbl,Daghigh:2023ixh,Shen:2023erj,Shen:2024qxv}.

%\subsection{Derivations}

More generally, Zhao \cite{Zhao:1995cp,Zhao:1996mr} shows that
double power-law density profiles of the form
\begin{equation} \label{dmprofilev2}
\rho(r)=\frac{\rho_{\rm ch}}{x^{a}\left(1+x^{1 / b}\right)^{b(g-a)}},
\end{equation}
%where $x = r/r_{\rm c}$,$r_{\rm c}$ a characteristic radius, and $\rho_{\rm ch}$ a characteristic density, 
have analytic expressions for the gravitational potential,
the enclosed mass, and the velocity dispersion (in terms of elementary functions) provided that $b = n$ and $g = 3 + k/n$, where $n$
and $k$ can be any natural numbers. The DZ profile is given by \cite{Freundlich_2020} \begin{equation} \label{dmprofilev3}
  \rho(r)=\frac{\rho_{\rm ch}}{x^{a}\left(1+x^{1 / 2}\right)^{2(3.5-a)}}.
\end{equation}

We use the Einstein cluster approach to model a stationary BH surrounded by a collection of gravitating masses. We refer the reader to \cite{Einstein:1939ms,Cardoso:2021wlq,Cardoso:2022whc,Maeda:2024tsg} for an extensive discussion, as well for technical details on the formalism, which has been originally applied to study binary BHs with an Hernquist-type matter distribution \cite{Hernquist:1990be}. 

In this framework the background metric specified by the line element 
\begin{equation}
ds^2=g_{\mu\nu}^{(0)}dx^\mu dx^\nu=-f(r) dt^2 + \frac{d r^2}{1-\frac{2m(r)}{r}} + r^2 d\Omega^2 \ ,
\end{equation}
is a solution of the sourced Einstein's field equations 
\begin{equation}
G_{\mu\nu}=8\pi T^{\rm env}_{\mu\nu}\ ,\label{math:fieldsback}
\end{equation}
where the properties of the environment are encoded by the anisotropic stress-energy  tensor with the following form:
\begin{equation}
    (T^{\rm env})^{\mu}{_\nu} = {\rm diag}(-\rho,0,P_t,P_t),
\end{equation}
\textcolor{black}{where $ \rho $ denotes the dark matter energy density and $ P_t $ encodes the isotropic tangential pressure arising from the orbital motion. This anisotropy is a defining characteristic of the Einstein cluster formalism, distinguishing it from perfect fluid descriptions and enabling the maintenance of hydrostatic equilibrium through centrifugal support rather than isotropic pressure gradients. Consequently, the dark matter halo remains stationary and self-consistent, with its gravitational influence encoded in the metric while preserving the condition $ P_r = 0 $ as a fundamental aspect of the model's stability and physical viability.} For a given choice of $\rho(r)$, the mass profile is determined by the continuity 
equation, while the metric variable $f(r)$ and the tangential 
pressure are determined by the $rr$-component of Eqs.~\eqref{math:fieldsback} and by the Bianchi identities, respectively.

\textcolor{black}{In this construction, we model a Schwarzschild black hole embedded within a spherically symmetric dark matter halo described by an Einstein cluster configuration. In such systems, dark matter is treated as a collection of non-interacting, collisionless particles confined to stable circular geodesics around the central mass. The assumption of purely circular motion eliminates any radial component of the particle velocities, thereby enforcing a strictly vanishing radial pressure ($ P_r = 0 $) throughout the halo. This condition is a direct consequence of the equilibrium dynamics of the system: with no radial infall or outflow, there is no radial momentum transfer, and thus, no mechanism to generate radial pressure. Moreover, this static, pressureless radial structure naturally implies that accretion of the halo onto the black hole is dynamically suppressed, with timescales for significant mass inflow exceeding the Hubble time, rendering the halo effectively stable over cosmological epochs.}

In the present coordinates, one can finally get three coupled equations: 
\begin{align}
    m'(r)&=4\pi Gr^2\rho(r), \lb{eqm} \\
    \frac{f'(r)}{f(r)}&=\frac{2m(r)}{r[r - 2m(r)]}, \lb{eqf} \\
    P(r)&=\frac{m(r)\rho(r)}{2[r-2m(r)]}. \lb{eqP}
\end{align}

Note that a prime denotes a derivative with respect to the radial coordinate. Note that it is supposed that lapse function in this form$
f(r) = \left(1 - \frac{2M_{\rm BH}}{r}\right) Y(r).
$ 
Eqs.~(\ref{eqm}) and (\ref{eqf}) can be solved by integration given suitable density profiles $\rho(r)$ given in Eq. \eqref{dmprofilev3}.

There are two cases for which mass $m(r)$ can calculated from Eq. \eqref{eqm}:
\begin{itemize}
    \item For $ x \ll 1 $ ($ r \ll r_{\rm c} $):
    \begin{equation}
        m(r) \approx \frac{4\pi \rho_{\rm ch} r^{3-a}}{(3-a) r_{}^{a-3}}.
    \end{equation}
    \item For $ x \gg 1 $ ($ r \gg r_{\rm c} $):
    \begin{equation}
        m(r) \approx -8\pi \rho_{\rm ch} r_{\rm c}^{2.5} r^{-0.5}.
    \end{equation}
\end{itemize}
To solve $f(r)$, we obtain the $Y(r)$ from \ref{eqP}:
\begin{itemize}
    \item For $r \ll r_{\rm c}$:
    \begin{equation}
        Y(r) =  \exp\left[\frac{2M_{\rm BH}}{r} + \frac{8\pi \rho_{\rm ch} r^{2-a}}{(3-a)(2-a) r_{\rm c}^{a-3}}\right].
    \end{equation}
    \item For $r \gg r_{\rm c}$:
    \begin{equation}
        Y(r) =  \exp\left(\frac{2M_{\rm BH}}{r} - \frac{32\pi \rho_{\rm ch} r_{\rm c}^{2.5}}{r^{0.5}}\right).
    \end{equation}
\end{itemize}
%Setting the boundary condition of asymptotic flatness, $f(r\rightarrow\infty)=1$,
Hence, we get the lapse functions as follows:
\begin{itemize}
    \item For $r \ll r_{\rm c}$, which we call Case 1:
    \begin{align} \label{sol1}
    f(r)=& \left(1 - \frac{2M_{\rm BH}}{r}\right) \\ \nonumber
    \times &\,\exp\left[\frac{2M_{\rm BH}}{r} + \frac{8\pi \rho_{\rm ch} r^{2-a}}{(3-a)(2-a) r_{\rm c}^{a-3}}\right],
    \end{align}
    \item For $r \gg r_{\rm c}$, which we call Case 2:
    \begin{equation} \label{sol2}
        f(r) = \left(1 - \frac{2M_{\rm BH}}{r}\right)  \exp\left(\frac{2M_{\rm BH}}{r} - \frac{32\pi \rho_{\rm ch} r_{\rm c}^{2.5}}{r^{0.5}}\right).
    \end{equation}
\end{itemize}
The event horizons for both cases are located at
\begin{equation} \label{horizon }
r_{\rm hor}=2M_{\rm BH}.
\end{equation} and there is a curvature singularity at $r = 0$. 

The metric functions $ f(r) $ provided in the two regimes, $ r \ll r_{\rm c} $ and $ r \gg r_{\rm c} $, smoothly reduce to the Schwarzschild solution in their respective limits. In the small $ r $ regime ($ r \ll r_{\rm c} $), the exponential correction term contains a dominant contribution proportional to $ r^{2-a} $. For $ a < 2 $, this term vanishes as $ r \to 0 $, leaving the metric function $ f(r) \to 1 - \frac{2M_{\rm BH}}{r} $, which is the Schwarzschild solution. Similarly, in the large $ r $ regime ($ r \gg r_{\rm c} $), the correction terms in the exponential decay as $ r^{-0.5} $ and $ r^{-1} $, becoming negligible as $ r \to \infty $. Consequently, the metric function again reduces to $ f(r) \to 1 - \frac{2M_{\rm BH}}{r} $. Therefore, in both asymptotic limits, the proposed metric naturally recovers the Schwarzschild solution, ensuring consistency with general relativity in vacuum.

To analyze the conditions under which the exponential corrections vanish, we solve the equations for $r$  in both regimes. For Case 1, we set
\begin{equation}
\frac{2M_{\rm BH}}{r} + \frac{8\pi \rho_{\rm ch} r^{2-a}}{(3-a)(2-a) r_{\rm c}^{a-3}} = 0,
\end{equation}
and we find the critical radius by solving for $r$ as
\begin{equation}
r_{\rm crit} = \left[\frac{8\pi \rho_{\rm ch}}{2M_{\rm BH} (3-a)(2-a) r_{\rm c}^{a-3}}\right]^{\frac{1}{a-3}}.
\end{equation}
This indicates that for small  $r$, the exponential correction term becomes negligible as  $r \to 0$, ensuring the metric approaches the Schwarzschild solution. Furthermore, the metric function at the critical radius becomes:
\begin{align}
f(r_{\rm crit, small}) =& 1 - \left(2M_{\rm BH}\right) \\ \nonumber
&\times \left[\frac{2M_{\rm BH} (3-a)(2-a) r_{\rm c}^{a-3}}{8\pi \rho_{\rm ch}}\right]^{\frac{1}{a-3}}.
\end{align}

\begin{figure}
    \centering
    \includegraphics[width=0.48\textwidth]{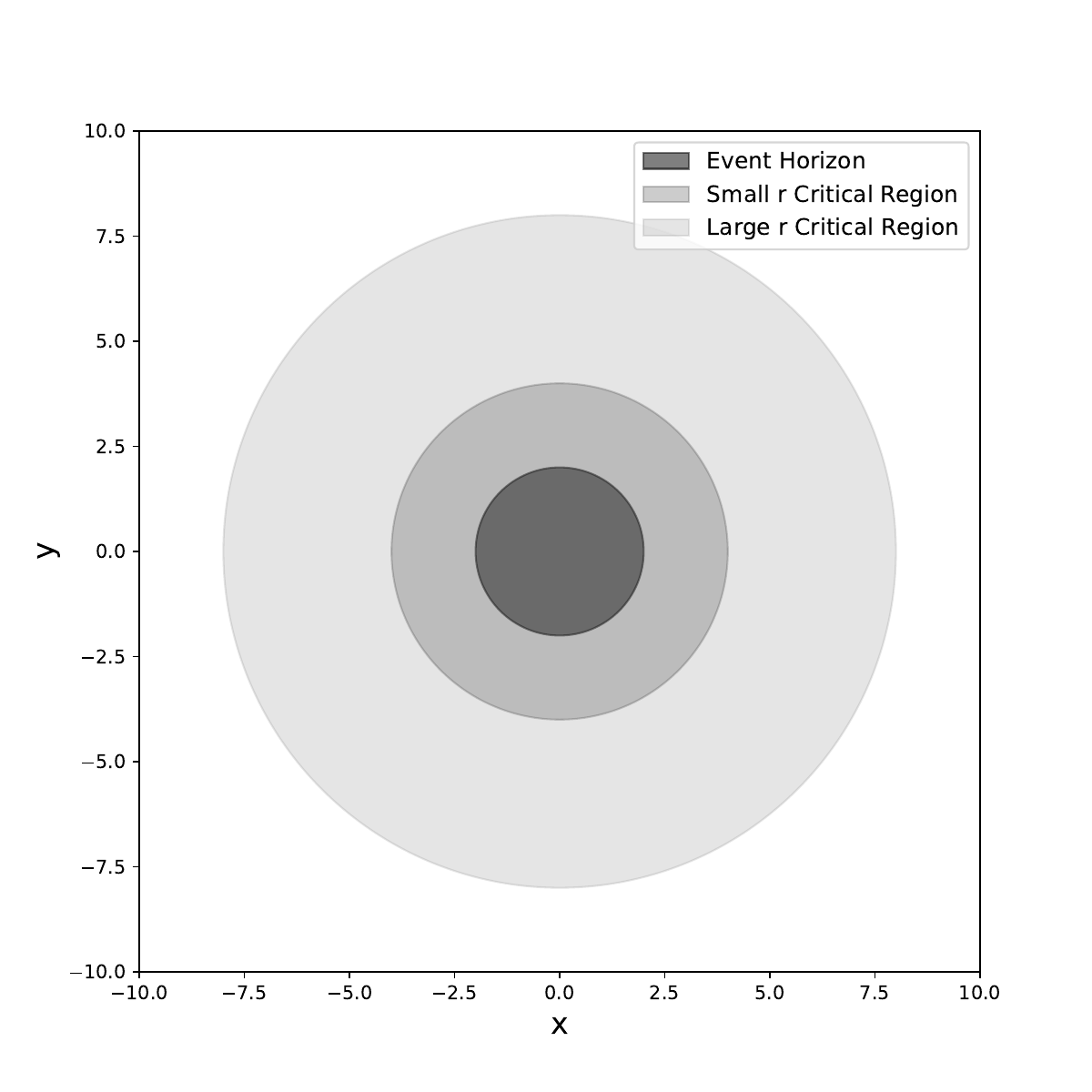}
    \caption{The figure shows the visualization of black hole, along with the critical radius $r_{\rm c}$.}
    \label{critical}
\end{figure}

For Case 2, we set
\begin{equation}
\frac{2M_{\rm BH}}{r} - \frac{32\pi \rho_{\rm ch} r_{\rm c}^{2.5}}{r^{0.5}} = 0,
\end{equation}
where we find the critical radius as
\begin{equation}
r_{\rm crit} = \frac{M_{\rm BH}^2}{0.000395786 \cdot r_{\rm c}^5 \rho_{\rm ch}^2}.
\end{equation}
This shows that for large  $r$ , the correction term in the exponential decays as  $r \to \infty$ , recovering the Schwarzschild solution. Also, the metric function at the critical radius becomes:
\begin{equation}
f(r_{\rm crit, large}) = 1 - \frac{0.000791572 \cdot r_{\rm c}^5 \rho_{\rm ch}^2}{M_{\rm BH}}.
\end{equation}
These expressions show how  f(r)  behaves at the critical radii in both regimes. See Fig. \ref{critical}.

The tangential pressure equals $ P_t = \rho/2 $ for any mass function. The anisotropic fluid "hair" surrounding the black hole satisfies both the weak and strong energy conditions everywhere because the pressure and density remain positive. The event horizon forms at $ r = 2m(r) $, where $ P_t/\rho $ diverges, violating the dominant energy condition near the horizon. Despite this violation, the region near the horizon contains almost no matter; the pressure and density become arbitrarily small in this area, so they do not influence the spacetime dynamics. 
\begin{figure}
    \centering
    \includegraphics[width=0.48\textwidth]{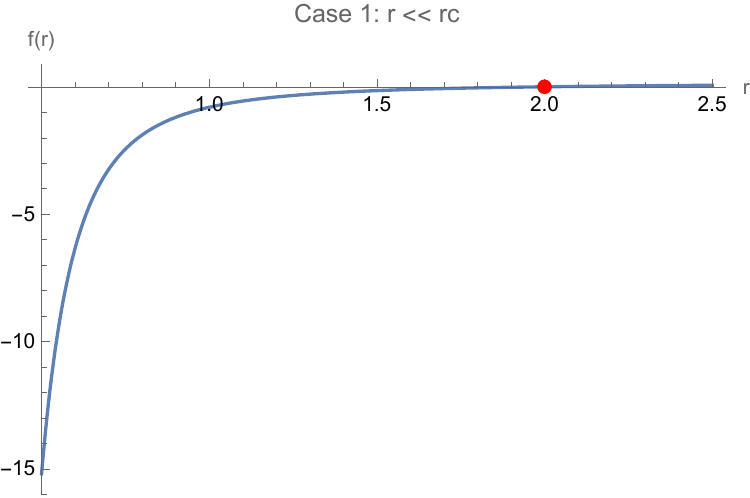}
    \includegraphics[width=0.48\textwidth]{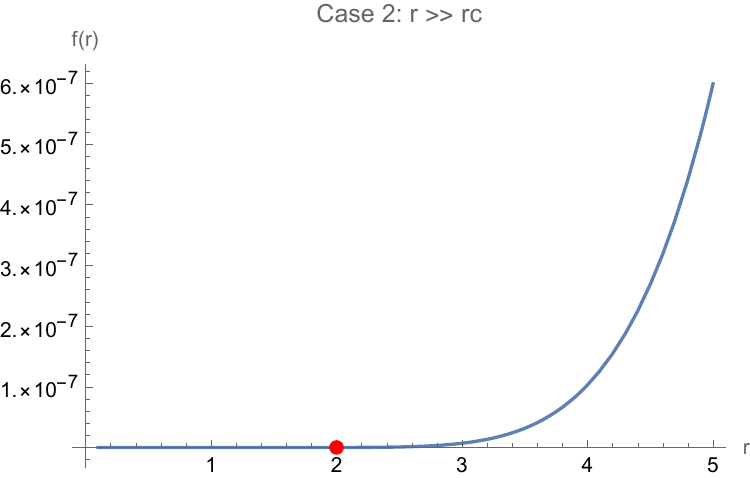}
    \caption{The \textcolor{black}{top} plot above shows the lapse function  $f(r)$  versus r for $r\gg r_{\rm c}$. The \textcolor{black}{bottom} plot above shows the lapse function  $f(r)$  versus r for $r \ll r_{\rm c}$.}
    \label{horizons}
\end{figure}

In Fig. \ref{horizons}, plot of $ f(r) $ versus $ r $ for $ r \ll r_{\rm c} $. The function $ f(r) $ incorporates the exponential growth due to the parameter $ \rho_{\rm ch} $ and power-law behavior $ r^{2-a} $. The red point indicates where $ f(r) $ crosses the x-axis, corresponding to the root of the equation $ f(r) = 0 $.

\section{Shadow and deflection angle of the DZ Black hole}  \label{sec3}
In this section, we study the shadow of the DZ black hole. First lets briefly explain the the effective potential which plays an important role in calculating the size of the black hole shadow.  The effective potential is defined as follows \cite{Vagnozzi:2019apd,Allahyari:2019jqz,Roy:2021uye,Afrin:2022ztr,Vagnozzi:2022moj,Paithankar:2023ofw}:
\begin{equation}
V_{\rm eff}(r) = \frac{f(r)}{r^2}. \label{veff}
\end{equation}
The asymptotic flatness of metric at spatial infinity implies that $V_{\rm eff}(r)$ falls as $1/r^2$ as $r \rightarrow \infty $ while at the outer horizon $r_H$ we have $V_{\rm eff}(r_H) = 0 $ since $f(r_H)=0$. Thus $V_{\rm eff}(r)$ has at least one maxima between $r_H$ and $ r \rightarrow \infty$. 

The relevant unstable extremum point $r_{\rm ph}$ described above leads to a circular orbit for photons and the value of the corresponding impact parameter, called \textcolor{black}{the critical impact parameter} $b_{\rm cr}$, then gives us the boundary of a black hole shadow. The black hole shadow for the observer at infinity,  leads to the critical impact parameter $r_{\rm sh} = b_{\rm cr}$. The critical impact parameter $b_{\rm cr}$ can be obtained  
\begin{equation}
b_{\rm cr} = \frac{1}{\sqrt{V_{\rm eff}(r_{\rm ph})}}.
\end{equation}
This corresponds to the turning point $r_{\rm ph}$ for which the effective potential is maximum. For the effective potential in Eq. (\ref{veff}), the photon sphere radius can be obtained as a solution to the equation \cite{Claudel:2000yi},
\begin{equation}
0 =\frac{d V_{\rm eff}}{dr}\Bigg|_{r=r_{\rm ph}} =\frac{r_{\rm ph}\, f'(r_{\rm ph})-2\, f(r_{\rm ph})}{r_{\rm ph}^3}.\label{photon sphere},
\end{equation}
where the prime denotes the derivative with respect to radial coordinate $r$. Using the radius of photon sphere $r_{\rm ph}$, it is then straightforward to get the critical impact parameter $b_{\rm cr}$ which gives the shadow radius $r_{\rm sh}$ of black hole as \cite{Vagnozzi:2022moj},
\begin{equation}
r_{\rm sh} = b_{\rm cr} = \frac{r_{\rm ph}}{\sqrt{f(r_{\rm ph})}},\label{shadow radius}
\end{equation}
where the last equality is obtained using the equation for the photon sphere radius Eq.(\ref{photon sphere}). Thus, given a specific form of $f(r)$, the size of the shadow of any black hole of the form of Eqs (\ref{sol1}) and (\ref{sol2})can be evaluated using Eq.(\ref{photon sphere}) and Eq.(\ref{shadow radius}). 

\textcolor{black}{The photon sphere is positioned at the radius defined by the equation $r = 3m(r)$ \cite{Cardoso:2008bp,Cardoso:2021wlq}, where the tangential pressure is given by $P = \rho/2$, applicable to any mass function. Therefore, the photon sphere $r_{\rm ph}$ for $r \ll r_{\mathrm{c}}$ is}

\begin{equation}
\boxed{
r_{\rm ph}
\;=\;
\biggl(\,\frac{3-a}{\,12\pi \rho_{\rm ch}\,}\biggr)^{\!\frac{1}{5-2a}}.
} \label{photonsphere}
\end{equation}

\textcolor{black}{On the other hand, in this large-\(r\) regime \(\bigl(r \gg r_{\mathrm{c}}\bigr)\), the approximate mass profile becomes negative, and the condition \(r = 3\,m(r)\) does not admit any real positive root.  Physically, this indicates that there is no real light ring in the far (\(r \gg r_{\mathrm{c}}\)) region under this approximation.}

 The radius of shadow to be equal to,
\begin{equation} \label{shadow of LLBH}
  R_{\mathrm{sh}}
  \;=\;
  \frac{r_{\mathrm{ph}}}{
    \sqrt{\Bigl(1 - \tfrac{2M_{\mathrm{BH}}}{r_{\mathrm{ph}}}\Bigr)\,
           \exp\!\Bigl[
             \tfrac{2M_{\mathrm{BH}}}{r_{\mathrm{ph}}}
             \;+\;
             \tfrac{8\pi\,\rho_{\mathrm{ch}}\,r_{\mathrm{ph}}^{\,2 - a}}
                   {(3-a)\,(2-a)\,r_{\mathrm{ch}}^{\,a - 3}}
           \Bigr]
    }
  }.
\end{equation}
\begin{figure}
    \centering
    \includegraphics[width=0.48\textwidth]{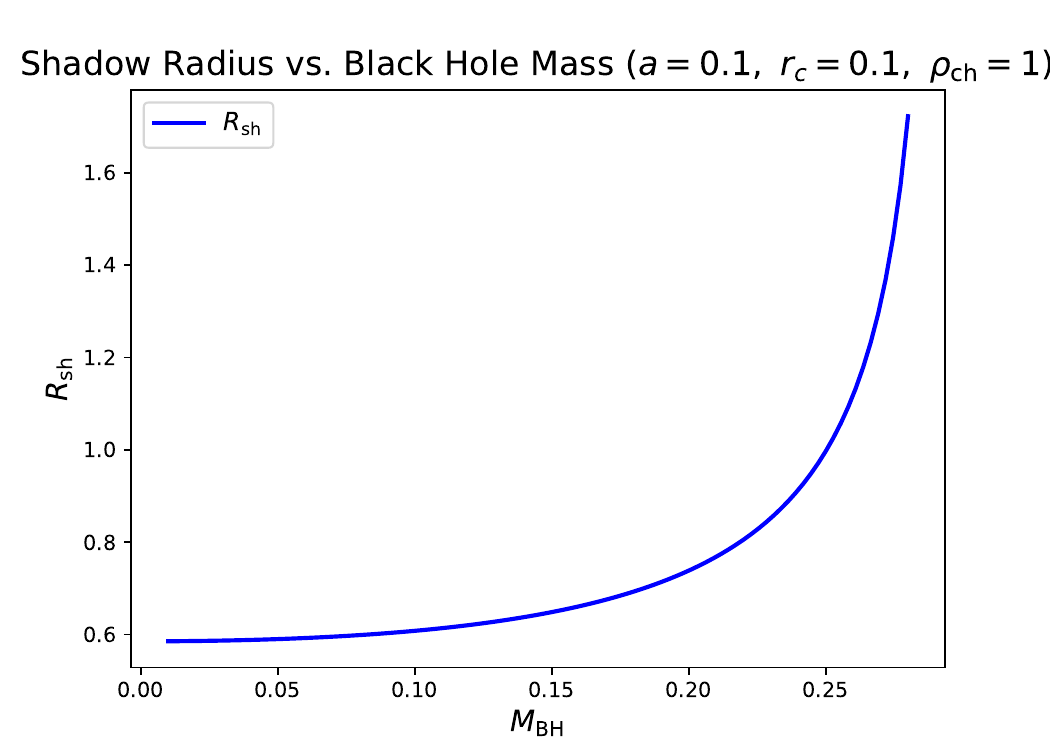}
\includegraphics[width=0.48\textwidth]{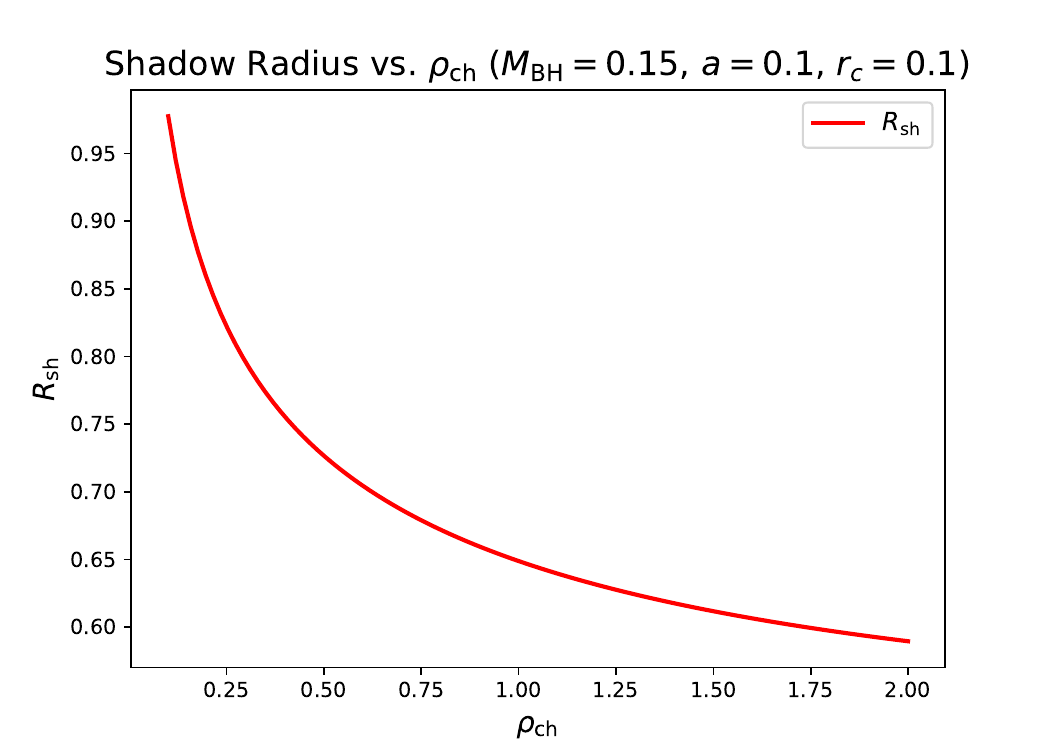}
    \caption{The above plot above shows the shadow radius  $r_{\rm sh}$  as a function of the black hole mass  $M_{\rm BH}$  for the two different  $Y(r)$  approximations. The below plot above shows the shadow radius  $r_{\rm sh}$  as a function of the central density  $\rho_{\rm ch}$. 
    }
    \label{shadow}
\end{figure}
\textcolor{black}{In Fig. \ref{shadow}, shows how the shadow radius varies with the black hole mass  $M_{\rm BH}$ and central density $\rho_{\rm ch}$.In this plot the shadow radius increases with increasing \(M_{\rm BH}\). In the top plot, for small \(M_{\rm BH}\), \(R_{\rm sh}\) starts at a relatively low value, but as \(M_{\rm BH}\) increases toward the upper limit of the domain (set by the requirement \(2M_{\rm BH}/r_{\rm ph} < 1\) so that the metric function remains positive), the shadow becomes significantly larger. This behavior is consistent with the notion that stronger gravitational lensing (i.e. a more massive black hole) results in a larger apparent shadow for a distant observer. The increase in \(R_{\rm sh}\) with \(M_{\rm BH}\) reflects the enhanced curvature of spacetime, leading to a greater deflection of light rays near the black hole. 
 In this bottom plot, the shadow radius decreases as the density \(\rho_{\rm ch}\) increases. For low values of \(\rho_{\rm ch}\), \(R_{\rm sh}\) is relatively larger, while a higher density yields a smaller shadow radius. This indicates that the presence of a denser matter field effectively “compresses” the effective photon–sphere, reducing the apparent size of the black‐hole shadow. The decrease in \(R_{\rm sh}\) with \(\rho_{\rm ch}\) suggests that the additional matter fields can significantly influence the observable properties of the black hole, in this case, by diminishing its shadow.} 

---

\begin{figure}
    \centering
    \includegraphics[width=0.48\textwidth]{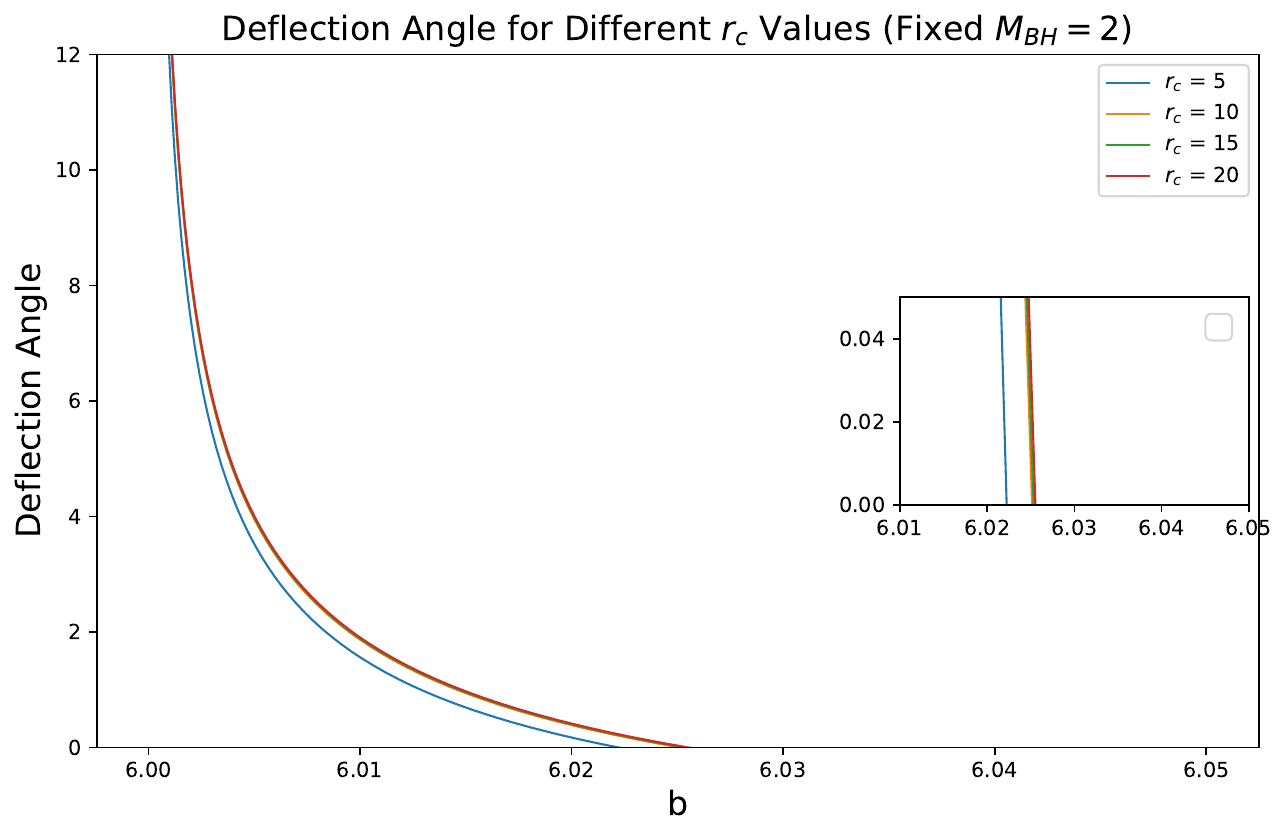}
    \caption{Deflection angle $\alpha$ as a function of the impact parameter $b$ for different characteristic radius $r_{\rm c}$, with $M_{\rm BH} = 2$. 
    The main plot illustrates the behavior of the deflection angle over the selected range, while the inset zooms in at $b \approx 6.01$ to highlight subtle differences between the curves for varying $r_{\rm c}$.}
    \label{deflection_angle}
\end{figure}

Lastly, we analyze the deflection angle numerically in weak field limits from the black hole described by the metric. For the metric , the bending angle in gravitational lensing is given by \cite{Virbhadra:1998dy,Virbhadra:1999nm}
\begin{equation}
    \alpha\left(r_{\rm ph}\right)=\int_{r_{\rm ph}}^{\infty} \frac{2}{\sqrt{B(r) C(r)}} \frac{\mathrm{d} r}{\sqrt{\frac{C(r)}{C\left(r_{\rm ph}\right)} \frac{A\left(r_{\rm ph}\right)}{A(r)}-1}}-\pi,
\end{equation}
where $r_{\rm ph}$ is the closest approach distance of the light ray to the black hole and $d s^2=-A(r) d t^2+\frac{d r^2}{B(r)}+C(r)\left(d \theta^2+\sin ^2 \theta d \phi^2\right)$. We plot the results in Fig. \ref{deflection_angle}.
We observe that the deflection angle $\alpha$ decreases as the critical radius $r_{\rm c}$ increases, reflecting the dependence of gravitational bending of light on the mass distribution characterized by $r_{\rm c}$. The inset provides a closer view around $b \approx 6.01$, revealing subtle differences between the curves for various $r_{\rm c}$ values. At smaller $r_{\rm c}$, the deflection angle tends to increase, indicating stronger gravitational effects. As for the general trend, the curves flatten at larger $b$, showing a gradual reduction in the deflection angle. This behavior aligns with the expectation that gravitational effects weaken as the impact parameter increases.

\section{Conclusions} \label{conc}
The study of black holes surrounded by the Dekel-Zhao (DZ) dark matter profile provides significant insights into how dark matter shapes the physical and observational properties of black holes. Using the approach by Cardoso et al., the researchers derived a modified Schwarzschild metric influenced by the DZ profile, revealing the critical interplay between dark matter density and black hole metrics. This analysis highlights the importance of parameters such as characteristic density ($\rho_{\rm ch}$), characteristic radius ($r_{\rm c}$), and profile exponent ($a$), as they dramatically affect observable phenomena, including the black hole’s shadow and gravitational lensing.

The metric’s behavior shows distinct characteristics in two regimes. In the small-radius regime ($r \ll r_{\rm c}$), the exponential corrections to the metric dominate, suggesting substantial deviations from Schwarzschild solutions near dense cores. Conversely, in the large-radius regime ($r \gg r_{\rm c}$), the corrections fade, restoring the Schwarzschild metric, thereby aligning with expectations of general relativity in vacuum.\textcolor{black}{The study of black hole shadows further reveals that the shadow radius grows exponentially with the black hole’s mass ($M_{\rm BH}$).} Notably, higher central densities, as encoded by DZ parameters, lead to smaller shadow radii, showcasing how dense environments modify the observable characteristics of black holes.

Photon dynamics within these configurations also underscore the influence of dark matter on gravitational phenomena. The photon sphere, critical impact parameters, and effective potential vary depending on the dark matter density profile, resulting in significant differences in shadow size and deflection behavior across different density regimes. The gravitational lensing effects are particularly sensitive to the characteristic radius ($r_{\rm c}$), with smaller radii yielding larger deflection angles. This aligns with expectations of stronger gravitational effects near more compact, dense cores, illustrating the nuanced relationship between dark matter profiles and light deflection near black holes.

In light of these findings, future research should focus on validating the theoretical predictions through detailed simulations and observational data, such as black hole shadow imaging from the Event Horizon Telescope. Expanding the framework to include rotating black holes and exploring the effects of broader ranges of DZ profile parameters could refine our understanding of dark matter's role in shaping black hole properties. Multi-messenger astronomy, integrating electromagnetic and gravitational wave observations, offers a promising avenue to constrain dark matter halo parameters. Additionally, extending this study to modified gravity theories and investigating the co-evolution of black holes and dark matter over cosmological timescales would provide deeper insights into the intricate dynamics of these systems.

\begin{comment}
The function $ Y(r) $ plays a crucial role in determining the shadow radius by modifying the effective lapse function $ f(r) $. Corrections to $ Y(r) $ from density parameters ($ \rho_{\rm ch} $, $ r_{\rm c} $, $ a $) significantly alter the size of the shadow radius. The shadow radius grows linearly with $ M_{\rm BH} $, as expected for spherically symmetric spacetimes. The small-$ r $ and large-$ r $ approximations converge for larger $ M_{\rm BH} $, where corrections become less impactful. Increasing $ \rho_{\rm ch} $ decreases the shadow radius, particularly in the large-$ r $ approximation. This implies that environments with high central densities could reduce the apparent size of the black hole shadow, which might be observable in astrophysical systems.  The small-$ r $ approximation dominates for compact cores and lower masses, where core corrections are significant, while the large-$ r $ approximation is better suited for diffuse or less dense environments.

These findings highlight the importance of density parameters and black hole mass in shaping the observable shadow radius. Observations of black hole shadows may help constrain these parameters in astrophysical systems.
\end{comment}

\section{Acknowledgments}
A. {\"O}. and R. P. would like to acknowledge networking support of the COST Action (CA) 18108 - Quantum gravity phenomenology in the multi-messenger appronkh, CA 22113 - Fundamental challenges in theoretical physics (Theory and Challenges), CA 21106 - COSMIC WISPers in the Dark Universe: Theory, astrophysics and experiments (CosmicWISPers), CA 21136 - Addressing observational tensions in cosmology with systematics and fundamental physics (CosmoVerse), and CA 23130 - Bridging high and low energies in search of quantum gravity (BridgeQG). We also thank TUBITAK and SCOAP3 for their support.
\bibliography{references.bib}

\end{document}